\documentclass{ws-ijmpa}

\begin{document}

\markboth{Hisakazu Minakata}{Long Baseline Neutrino Experiments with Two-Detector Setup}

%
\catchline{}{}{}{}{}
%

\title{Long Baseline Neutrino Experiments with Two-Detector Setup}

\author{HISAKAZU MINAKATA}

\address{Department of Physics, Tokyo Metropolitan University, 
Hachioji, Tokyo 192-0397, Japan
\\
minakata@phys.metro-u.ac.jp}

\maketitle


\begin{abstract}
I discuss why and how powerful is the two-detector setting in 
neutrino oscillation experiments. I cover three concrete examples: 
(1) reactor $\theta_{13}$ experiments, 
(2) T2KK, Tokai-to-Kamioka-Korea two-detector complex for 
measuring CP violation, determining the neutrino mass hierarchy, 
and resolving the eight-fold parameter degeneracy, 
(3) two-detector setting in a neutrino factory at baselines 
3000 km and 7000 km for detecting effects of non-standard interactions 
(NSI) of neutrinos.

\end{abstract}


\section{Introduction}

Unified understanding of the physics of quark and lepton flavor mixings 
would be the most important goal for the contemporary flavor physics.
Though we started to grasp the structure of the flavor mixing matrix, 
the MNS matrix\cite{MNS}, there is a long way to go. 
Unlike the quark sector in which the dominant mechanism of 
CP violation is identified\cite{KM}, the very existence of CP violation 
itself remains a mystery in the lepton sector.
Therefore, looking for leptonic CP violation will be one 
of the crucial key elements in planning the next generation 
neutrino experiments. 
Moreover, various studies indicated that uncovering leptonic 
CP violation is highly challenging experimentally.  
Therefore, strategic thoughts on how to make the goal may be 
of some use. 
This is the only reason I can think of why this talk with such a 
technical title (though it was given by the organizer) may be worth to 
be presented in the flavor physics conference.

Yet, I will try to cover the related topics in a slightly wider context 
under the hope that it serves for  illuminating the merits of the 
two-detector setting even more clearly. 
Namely, I address the three concrete examples of the two-detector setting;\footnote{
We define the two-detector setting as composed of two detectors 
excluding a front detector which measures un-oscillated neutrino flux or  monitors beam. 
}

\begin{itemlist}

\item

Reactor $\theta_{13}$ experiments\cite{krasnoyarsk,MSYIS}

\item

T2KK, Tokai-to-Kamioka-Korea identical two-detector complex for 
measuring CP violation, determining the neutrino mass hierarchy, 
and resolving the eight-fold parameter degeneracy\cite{T2KK1st,T2KK2nd}

\item

Two-detector setting in a neutrino factory (3000 km, 7000 km) 
for detecting non-standard interactions (NSI) of neutrinos\cite{NSI-nufact}

\end{itemlist}

\noindent
Before entering into the discussions let us raise a general question; 
What is good in two-detector setting?
The answer is:

\begin{itemlist}

\item

The systematic errors cancel between the two detectors. 

\item

Measurement at the two detectors can have synergy effects 
whose significance, however, varies a lot in case by case.

\end{itemlist}

\vspace{-0.25cm}
\section{Reactor $\theta_{13}$ Experiments}

With regard to multi detector reactor experiment, there is in fact, 
an ancestor experiment, the Bugey experiment\cite{bugey} which 
utilized the three detectors albeit not quite identical ones. 
It was proposed in \cite{krasnoyarsk,MSYIS} that the only practical 
way to measure a small depletion due to $\theta_{13}$ is to place 
two identical detectors one at a near (100-300 m) and the other 
at a far (1-2 km) locations. 
Controlling the systematic errors and cancelation of them between 
the two detectors is the key to such difficult measurement.
Now it becomes a ``customary'' design for the reactor $\theta_{13}$ 
experiments and the principle is employed in all the projects in 
construction\cite{3projects}. 
See \cite {reactor_white} for other projects.

\vspace{-0.25cm}
\section{T2KK;  Tokai-to-Kamioka-Korea Two Detector Complex}

In the context of accelerator neutrino experiments a proposal 
of two detector setting appeared in the Brookhaven proposal\cite{brookhaven}. 
The authors of Ref.~\cite{MNplb97} discussed two detector methods 
for measuring leptonic CP violation by observing neutrino oscillation 
``phase'' at two different locations. 
A concrete realization of this principle was proposed 
\cite{T2KK1st,T2KK2nd} in a form of identical two-detector 
setting using two megaton class detectors in Kamioka and Korea 
receiving an intense neutrino beam from J-PARC, 
the Tokai-to-Kamioka-Korea (T2KK) project. 
See \cite{T2KKweb} for more about the project.

\begin{figure}[htb]
\begin{center}
\epsfig{file=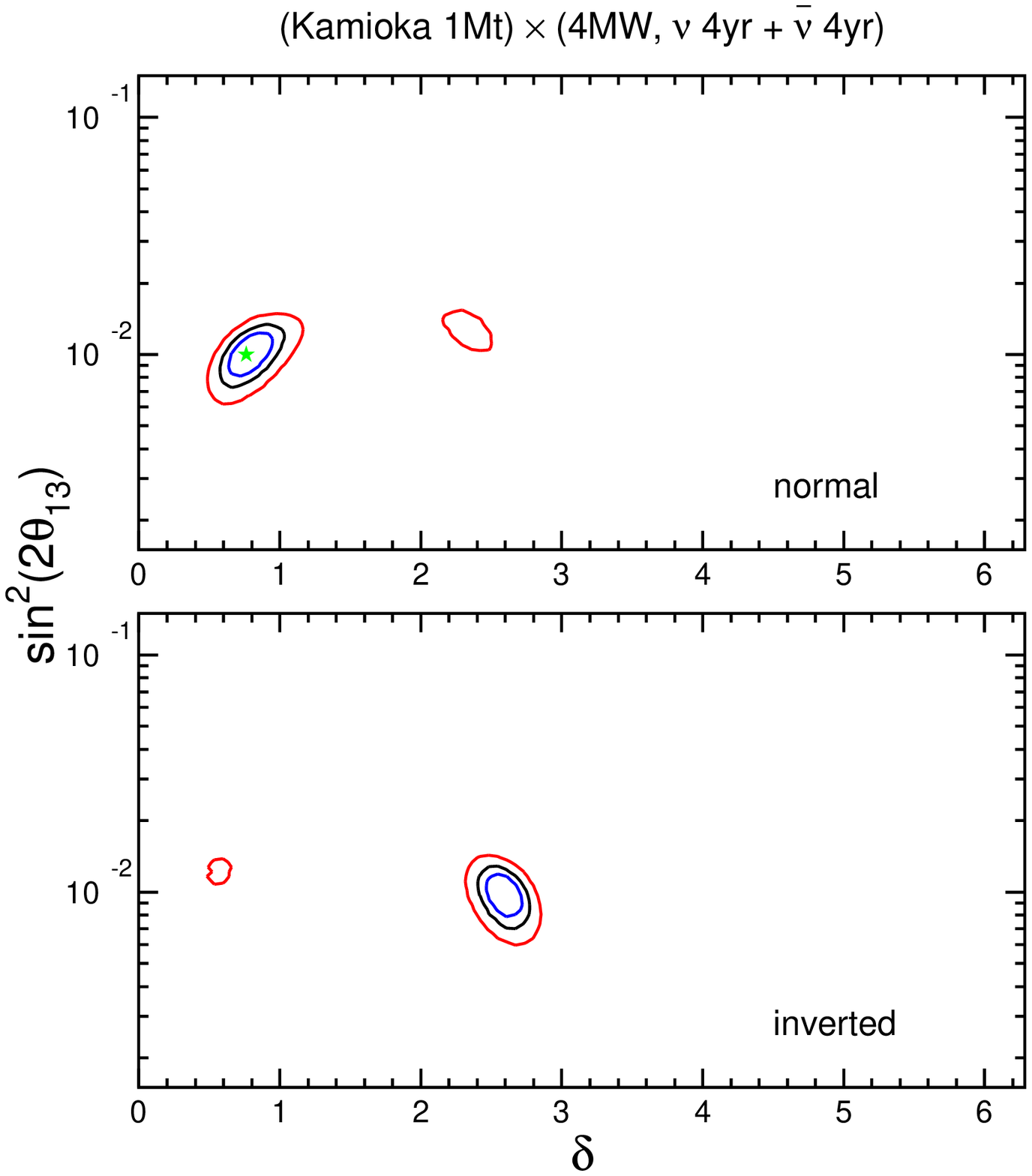,height=2.8 in}
\epsfig{file=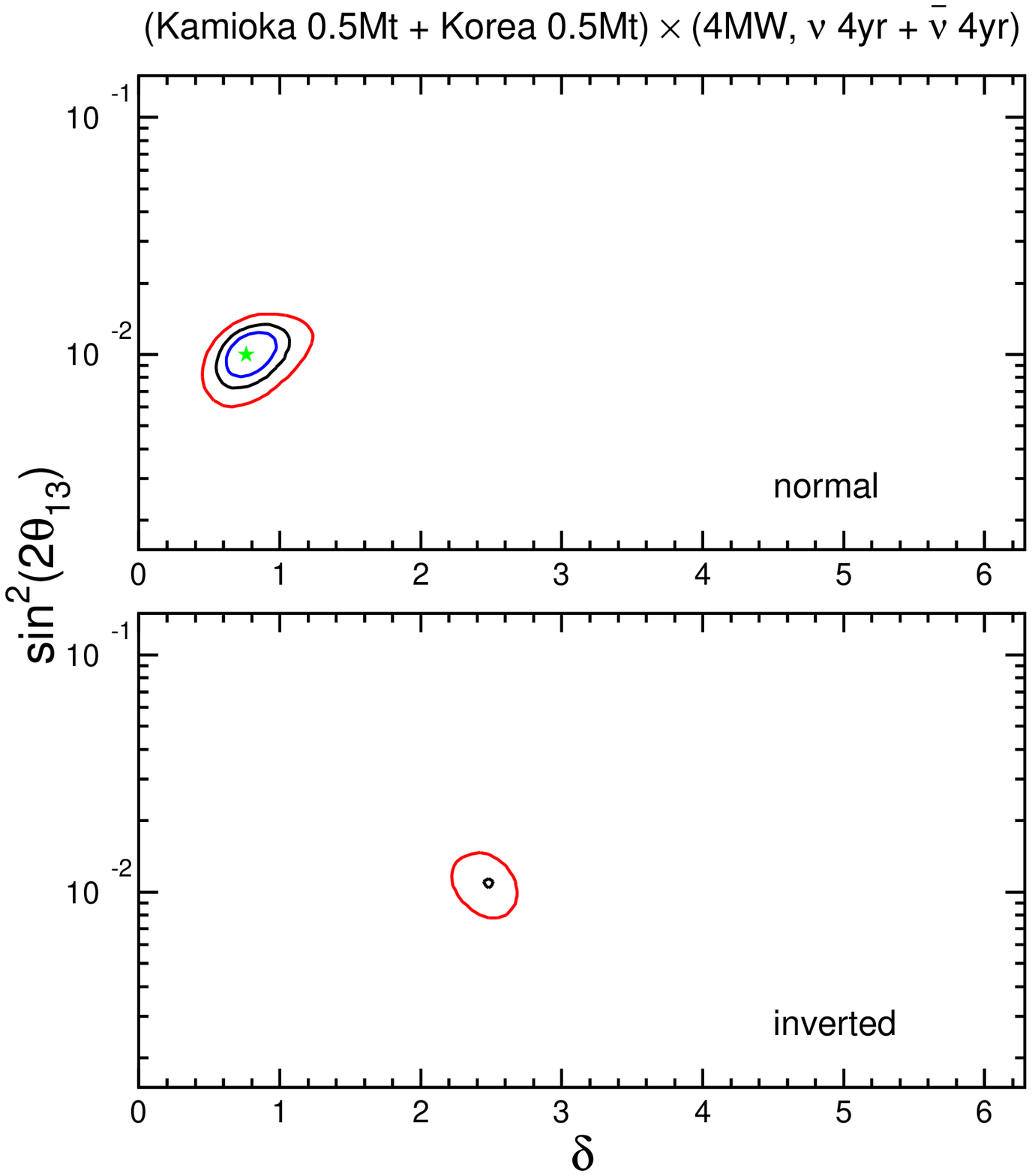,height=2.8 in}
\caption{The region allowed in $\delta-\sin^2 2\theta_{13}$ space by 4 years of 
neutrino and antineutrino running in  T2K II (left panels), and the 
Kamioka-Korea two detector setting (right panels). 
They are taken from the supplementary figures behind the 
reference\protect\cite{T2KK1st} to which the readers are referred 
for details of the analysis. 
Notice that the standard setting in T2K II, 2 (6) years of 
neutrino (antineutrino) running, leads to a very similar results 
(as given in \protect\cite{NOVE06_mina}) to the one presented in the left 
panel of this figure.  
The true solutions are assumed to be located at 
($\sin^2{2\theta_{13}}$ and $\delta$) = (0.01, $\pi/4$) 
with positive sign of $\Delta m_{31}^2$, as indicated as the green star. 
The intrinsic and the $\Delta m_{31}^2$-sign clones appear in the 
same and the opposite sign $\Delta m_{31}^2$ panels, respectively. 
Three contours in each figure correspond to
the 68\% (blue line), 90\% (black line) and 99\% 
(red line) C.L. sensitivities, respectively.
}
\label{intrinsicKamKorea}
\end{center}
\end{figure}

I just give a sketchy description here about how the T2KK two 
detector setting is powerful. 
For more details, in particular, for a fuller description of the 
sensitivities to CP violation,  the mass hierarchy, and resolution 
of the eight-fold parameter degeneracy \cite{intrinsic,MNjhep01,octant}, 
see \cite{T2KK1st,T2KK2nd}. 
Figure~\ref{intrinsicKamKorea} shows how the spectrum information 
is powerful to determine CP phase $\delta$ resolving the 
$\delta \leftrightarrow \pi - \delta$ ambiguity.
Comparison between the left and the right panels indicates 
that the T2KK setting is more efficient to resolve the ambiguity 
by comparing the yields at the two detectors at the two different locations.

\begin{figure}[htb]
\begin{center}
\epsfig{file=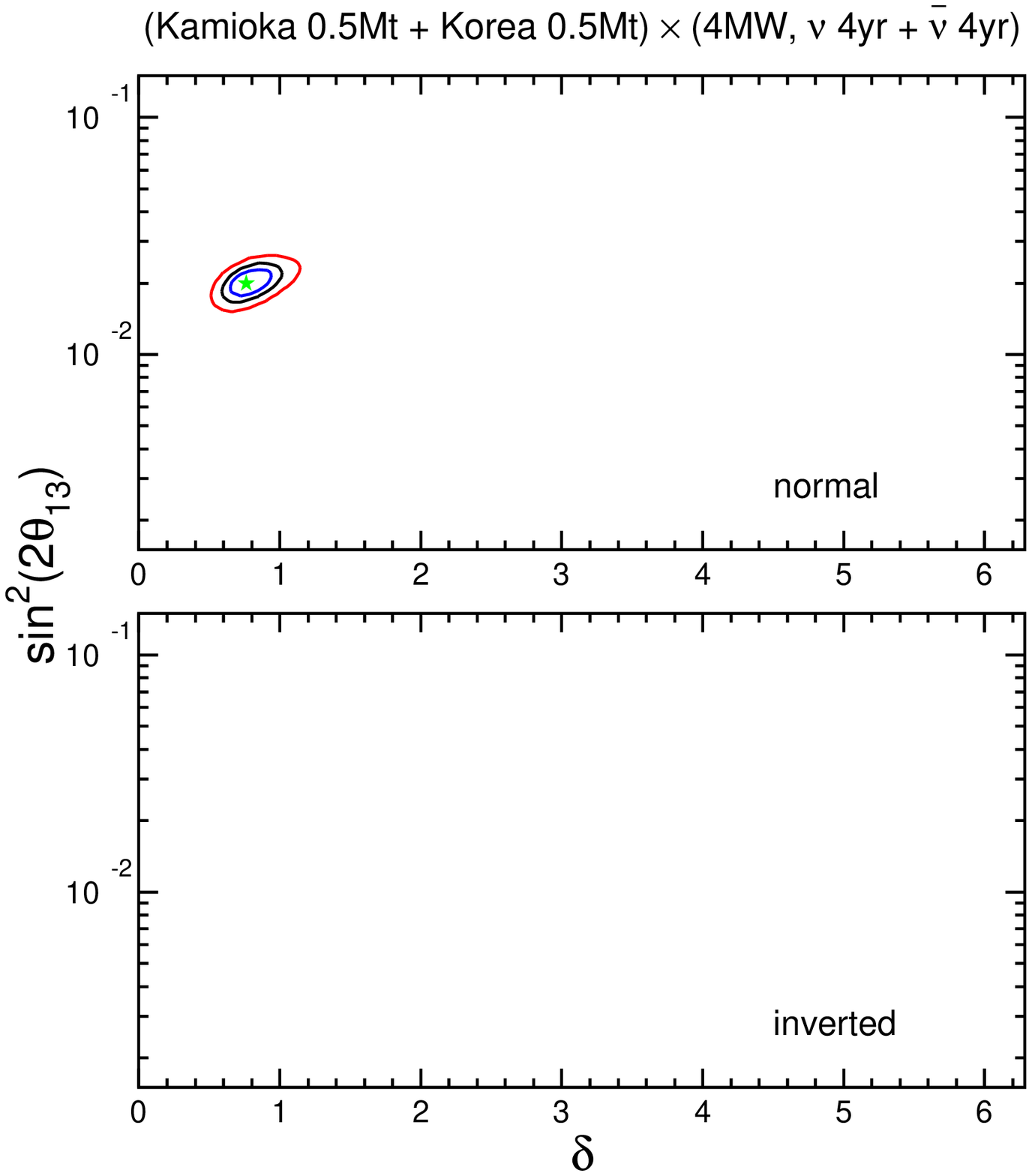,height=2.8 in}
\epsfig{file=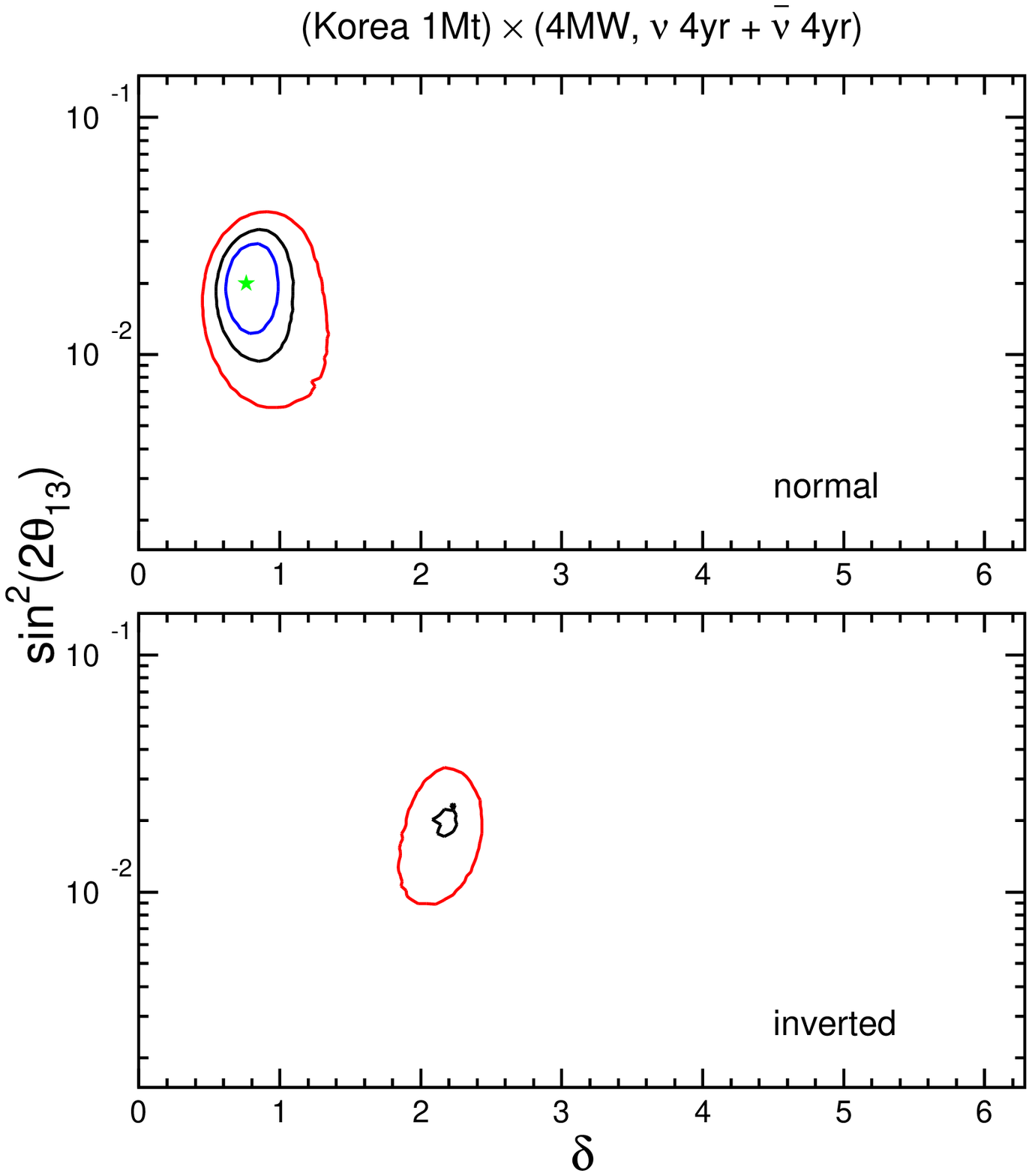,height=2.8 in}
\caption{The similar sensitivity plot as in Fig.~\ref{intrinsicKamKorea}. 
The left panels are for T2KK and the right panels are for a single 
0.54 megaton detector placed in Korea.
}
\label{signdm2KamKorea}
\end{center}
\end{figure}

It is often said that resolution of the mass hierarchy can be done 
by using long baseline thanks to the earth matter effect to neutrino 
oscillation. Though it is of course true, Fig.~\ref{signdm2KamKorea} 
indicates that it is {\em not} the whole story. 
The left panels are for the T2KK setting with each 0.27 Mton fiducial mass 
detectors placed in Kamioka and Korea, whereas the right panels 
are for Korea only setting with 0.54 Mton fiducial mass. 
The figure demonstrates that the two detector comparison has a higher 
resolving power of the neutrino mass hierarchy than the Korea only setting. 
Though I do not elaborate, resolution of the $\theta_{23}$ octant 
degeneracy is also merited by the two detector setting which does (Korea) 
and does not (Kamioka) feel the solar oscillation effect. 
There is an interesting competition and synergy between the 
T2KK and the reactor-acelerator combined method \cite{MSYIS,hiraide} 
for lifting the $\theta_{23}$ degeneracy. 
The former (latter) is more powerful at small (large) $\theta_{13}$.

\vspace{-0.25cm}
\section{Probing Non-Standard Neutrino Interactions at Neutrino Factories}

My last topics is the two detector setting in neutrino factory, 
one at $\sim$3000 km and the other at $\sim$7000 km, the latter 
so called the magic baseline\cite{huber-winter}. 
The idea of the setting was originated from the consideration 
of how the intrinsic $\theta_{13} - \delta$ degeneracy can 
be lifted\cite{intrinsic,huber-winter}. 
It has been also shown that a detector at the magic baseline 
has an extremely high sensitivity for measuring the average 
earth matter density (assuming the MSW theory) 
along the neutrino trajectory\cite{mina-uchi,gandhi-winter}.

Therefore, it is entirely natural to think about the possibility that neutrino 
factory with two detector setting can serve for a powerful hunting 
tool for possible non-standard interactions 
(NSI)\cite{wolfenstein,grossmann} possessed by neutrinos. 
It is conceivable that such NSI would arise if there exists 
new physics scale at TeV ranges. 
They may be parametrized by four Fermi interactions;  
${\cal L}_{\text{eff}}^{\text{NSI}} =
-2\sqrt{2}\, \varepsilon_{\alpha\beta}^{fP} G_F
(\overline{\nu}_\alpha \gamma_\mu P_L \nu_\beta)\,
(\overline{f} \gamma^\mu P f)$.
One expects by dimensional counting that NSI coefficients 
$\varepsilon_{\alpha\beta}^{fP}$ and therefore the effective interaction 
$\varepsilon_{\alpha\beta} \equiv \sum_{f,P} \frac{n_f}{n_e}
\varepsilon_{\alpha\beta}^{fP}$ which appears in the neutrino 
evolution equation would have a size of the order of 
$(m_{Z} / M_{np})^2 \sim 10^{-2}$ ($10^{-4}$) for  
$M_{np} = 1$ (10) TeV. 
Then, neutrino factory is the best thinkable machine to explore 
such tiny effects of NSI. 
There exist numerous references which are devoted to existing 
constraints on NSI and how to probe it further by future experiments. 
See e.g., the references quoted in \cite{NSI-nufact}.

\begin{figure}[htb]
\begin{center}
\vspace{-0.4cm}
\epsfig{file=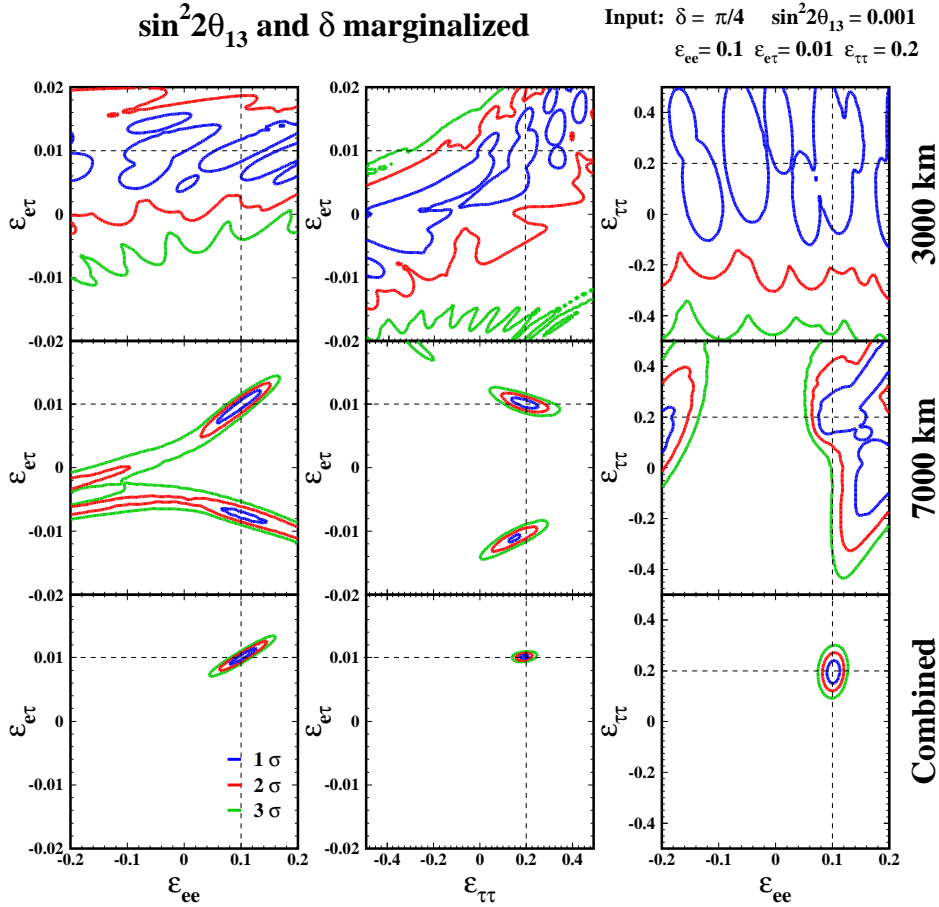,height=5.0 in}
\caption{Allowed regions projected into the plane of 
2 NSI parameters, $\varepsilon_{ee}$-$\varepsilon_{e\tau}$
(left panels), $\varepsilon_{\tau\tau}$-$\varepsilon_{e\tau}$ (middle panels)
and $\varepsilon_{ee}$-$\varepsilon_{\tau\tau}$ (right panels) 
corresponding to the case where 
the input parameters are as follows: 
$\sin^2 2\theta_{13} = 0.001$, 
$\delta = \pi/4$, 
$\varepsilon_{e\tau} = 0.01$, 
$\varepsilon_{ee} = 0.1$, and 
$\varepsilon_{\tau\tau} = 0.2$. 
The neutrino energy is $E_\mu$ = 50 GeV and the baseline is taken as 
$L=3000$ km (upper panels), 7000 km (middle horizontal panels) 
and combination (lower pannels). 
The thin dashed lines are to indicate the input values of 
$\varepsilon_{\alpha \beta}$.
The fit was performed by varying freely 4 parameters, 
$\theta_{13}$, $\delta$ and 2 $\varepsilon$'s 
with $\theta_{13}$ and $\delta$ being marginalized. 
The number of muons decays per year is $10^{21}$, 
the exposure considered is 4 (4) years for neutrino (anti-neutrino),
and each detector mass is assumed to be 50 kt. 
Notice that this figure supplements Fig.~16 of \protect\cite{NSI-nufact} 
which uses the same parameters as this figure except for the input 
value of CP phase, $\delta=3\pi/2$. 
}
\label{ee-et-tt-piby4-nzero}
\end{center}
\vspace{-0.4cm}
\end{figure}

Now, I present in Fig.~\ref{ee-et-tt-piby4-nzero} the regions allowed by 
measurement of detectors at 3000 km (upper panels), 
7000 km (middle panels), and two detectors combined (bottom panels). 
The left, middle, and the right three panels are for the cases with 
$\varepsilon_{ee}$-$\varepsilon_{e\tau}$, 
$\varepsilon_{\tau\tau}$-$\varepsilon_{e\tau}$, and 
and $\varepsilon_{ee}$-$\varepsilon_{\tau\tau}$, 
respectively.  
%
%
We notice that the detector at 3000 km alone does not have 
good sensitivities to NSI. 
This statement is also true for the detector at 7000 km though 
the sensitivity to the off diagonal elements is much better than 
that of 3000 km detector; It {\em is} the very motivation for 
placing the detector at the magic baseline. 
The reason for such disparity in the sensitivities to the diagonal and 
the off diagonal elements of $\varepsilon_{\alpha\beta}$ is explained 
in  \cite{NSI-nufact}  based on the analytic formula derived there. 
The synergy effect of combining the intermediate and the far 
detectors is remarkable. 
The allowed regions scattered in wide ranges in the top (3000 km) and 
the middle (7000 km) panels combine into a much smaller region in 
the bottom panel. 
To the best of my knowledge, such a synergy effect so significant 
as in Fig.~\ref{ee-et-tt-piby4-nzero} is rarely seen.

The remaining (important!) issue in the neutrino factory measurement 
of NSI is that the sensitivity to $\theta_{13}$ is largely lost because 
of confusion with NSI \cite{confusion}. 
We were able to show that this problem is also solved by the 
same two detector setting. 
See Ref.~\cite{NSI-nufact} for further details. 
Therefore, we have concluded (I believe for the first time) 
that the results obtained in this paper open the door to 
the possibility of using 
{\em neutrino factory as a discovery machine for NSI}, 
while keeping its function of precision measurement 
of lepton mixing parameters. 
Finally, 
I would like to emphasize that discovery of physics beyond 
the neutrino mass incorporated Standard Model would be 
much more exciting goal for remote future neutrino experiments.

\section*{Acknowledgments}

I would like to thank the organizers of ICFP 2007 for the invitation. 
My talk is based on works done by fruitful collaborations with my friends, 
in particular, Takaaki Kajita, Hiroshi Nunokawa, Renata Zukanovich Funchal. 
This work was supported in part by KAKENHI, Grant-in-Aid for
  Scientific Research, No 19340062, Japan Society for the Promotion of
  Science,


\end{document}